\documentclass[12pt]{article}
\usepackage{graphicx}
\usepackage{amsmath}
\usepackage{amssymb}
\usepackage{caption2}
\begin{document}
\bibliographystyle{prsty}
\begin{center}
{\large {\bf \sc{  Magnetic moment of the pentaquark $\Theta^+(1540)$ with QCD sum rules }}} \\[2mm]
Zhi-Gang Wang$^{1}$ \footnote{Corresponding author; E-mail,wangzgyiti@yahoo.com.cn.  }, Wei-Min Yang$^{2}$ and Shao-Long Wan$^{2} $    \\
$^{1}$ Department of Physics, North China Electric Power University, Baoding 071003, P. R. China \\
$^{2}$ Department of Modern Physics, University of Science and Technology of China, Hefei 230026, P. R. China \\
\end{center}

\begin{abstract}
In this article, we study the magnetic moment of the pentaquark state $ \Theta^+(1540)$
with the QCD sum rules approach in the external electromagnetic field. The numerical results
indicate the magnetic moment of the pentaquark state $ \Theta^+(1540)$ is about $\mu_{\Theta^+}=(0.24\pm0.02)\mu_N$.
\end{abstract}


\section{Introduction}

Several collaborations have reported the observation of the
new baryon state $\Theta^+(1540)$ with positive strangeness
and minimal quark contents $udud\bar{s}$ \cite{exp2003}. The
existence of such an exotic state with narrow width $\Gamma < 15
MeV$ and $J^P={\frac{1}{2}}^+$ was first predicted by Diakonov,
Petrov and Polyakov  in the chiral quark soliton model, where the
$\Theta^+(1540)$ is a member of the baryon antidecuplet $\overline {10}$
\cite{Diakonov97}. The discovery   has
opened a new field of strong interaction  and provides a new
opportunity for a deeper understanding of the low energy QCD. Intense
theoretical investigations have been motivated
 to clarify the quantum numbers and to understand the
under-structures of the pentaquark state $\Theta^+(1540)$
\cite{jaffe03,karliner03,zhu03,matheus04,sugiyama04,eide04,Wang05,Narison04,
Kanada04,takeuchi04,LatP,LatN,Latt04,Shuryak04}. The zero of the third component of isospin
$I_3=0$ and the absence of isospin partners suggest  that the pentaquark state
$\Theta^+(1540)$ is an isosinglet, while the spin and parity have
not been experimentally determined yet and  no consensus has ever
been reached  on the theoretical side. Unlike the chiral soliton model, the
 quark models treat the constituent quarks or quark clusters as the elementary
degrees of freedom,  there exist a great number of  possible quark configurations satisfy the Fermi
statistics and the color singlet condition
 for the substructures of the pentaquark state $\Theta^+(1540)$
  if we release  stringent dynamical constraints.
  Different models can lead to
different predictions, for example, the chiral soliton
model \cite{Diakonov97}, the diquark-diquark-antiquark  model
\cite{jaffe03}, the triquark-diquark model \cite{karliner03}, some
QCD sum rules approaches \cite{matheus04}, some lattice QCD \cite{LatP} and some
quark  models (or quark potential models)
\cite{riska03,hosaka03,Carlson03p} prefer positive parity,  while
other QCD sum rules approaches \cite{zhu03, sugiyama04,Wang05}, some lattice QCD \cite{LatN} and
some quark  models \cite{zhang03,carlson03n,wu03} favor
negative parity, we must select
the preferred configurations with the elementary quantum numbers.

 Determining the parity of the pentaquark state  $\Theta^+(1540)$
 is of great importance in establishing its basic quantum
numbers and in understanding the low energy QCD especially when
multiquarks are involved.
The experiments of photo- or electro-production and proton-proton collision can be used
  to determine the fundamental quantum numbers of  the pentaquark state $\Theta^+(1540)$, such
  as spin and parity \cite{HosakaPP,Nam04M}.
  In fact, the magnetic moment of the $\Theta^+(1540)$ $\mu_{\Theta^+}$ is an important ingredient
in studying the cross sections of the photo-production,  and may be extracted from
the experiments eventually in the near future. The magnetic moments of the pentaquark states are
  fundamental parameters as their  masses, which have  copious  information  about the underlying quark
structures, can be used to
distinguish the preferred quark configurations from  various  theoretical models and  deepen our
understanding of the underlying dynamics.

 There have been several works on the magnetic moment $\mu_{\Theta^+}$
  \cite{Nam04M,Zhao04M,Kim04M,Goeke04M,Zhu04M,Bijker04M,Inoue04M,Hong04M,Delgado04M,Huang04M},
in this article, we take the point of view that the quantum
numbers of the pentaquark state  $\Theta^{+}(1540)$ are $J=\frac{1}{2}$ ,
$I=0$ , $S=+1$, and study its magnetic moment $\mu_{\Theta^+}$ with the QCD sum rules approach \cite{Shifman79}.

 The article is arranged as follows:   we derive the QCD sum rules in the external
 electromagnetic field for the magnetic moment $\mu_{\Theta^+}$ in section II;
 in section III, numerical results; section IV is
reserved for conclusion.

\section{QCD Sum Rules in External Electromagnetic Field}
In the QCD sum rules approach, the operator product expansion
 is used to expand the time ordered interpolating currents into a series of quark and
 gluon condensates which parameterize the long distance properties, while
 the short distance effects are incorporated in the Wilson coefficients.
 Although for medium and asymptotic momentum transfers
 the operator product expansion method can be applied for the form factors and  moments of wave
 functions \cite{Chernyak84}, at low momentum transfer,
 the standard operator product expansion method cannot be  consistently  applied,
 as pointed out in the early work on photon
 couplings at low momentum for the nucleon  magnetic moments
 \cite{Ioffe84,Balitsky83}. In Refs.\cite{Ioffe84,Balitsky83}, the problem was solved by using a
 two-point correlation function in an external electromagnetic field, with
  vacuum susceptibilities introduced as parameters for
 nonperturbative propagation in the external field, i.e. the QCD  sum rules in the external field.
 As  nonperturbative vacuum properties, the vacuum susceptibilities can be
 introduced for both small and large momentum transfers  in the
 external fields .

In the following, we write down  the two-point correlation function $\Pi_{\eta}(p)$ in the presence of a weak external
electromagnetic field $F_{\alpha\beta}$,
\begin{eqnarray}
\Pi_{\eta}(p)&=& i\int d^4 x e^{ip\cdot x}\langle 0|T \{ \eta(x){\bar \eta}(0) \}|0\rangle_{F_{\alpha\beta
} } \, , \nonumber \\
&=&\Pi_0(p) + \Pi_{\mu\nu} (p) F^{\mu\nu} +\cdots \, ,
\end{eqnarray}
where the $\Pi_0(p)$ is the correlation function  without the external field
$F_{\alpha\beta}$ and the $\Pi_{\mu\nu} (p)$ is the linear response term, the $\eta(x)$ is the interpolating current with the quantum numbers of the pentaquark state $\Theta^+(1540)$ \cite{zhu03}
\footnote{Here we use the diquark-triquark type interpolating current suggested  in Ref.\cite{zhu03},
the sum rules with a diquark-diquark-antiquark type interpolating current are presented in Refs.\cite{Wang052,Wang053},
 the sum rules with  other interpolating currents will be our next work. Comparing
with the magnetic moments obtained from different approaches and detailed studies
may shed light on the under-structures and low energy dynamics of
the pentaquark states.  },
\begin{eqnarray}
\eta(x)={1\over \sqrt{2}} \epsilon^{abc} \left\{u^T_a(x) C\gamma_5
d_b (x)\right\} \{ u_e (x) {\bar s}_e (x) i\gamma_5 d_c(x) - d_e (x) {\bar s}_e (x) i\gamma_5 u_c(x)  \} \, ,
\end{eqnarray}
and
\begin{equation}
\langle 0| \eta(0) |\Theta^+ (p)\rangle =f_0 u(p) \, ,
\end{equation}
here the $a$, $b$, $c$ and $e$ are color indexes. The interpolating current in Eq.(2) can couple to the pentaquark state
with both negative and positive parity, and picks out only the state with lowest
mass without knowledge about its parity. As the electromagnetic vertex of the pentaquark states with
either negative or positive parity is the same, we can extract the
absolute value of the magnetic moment of the lowest pentaquark state.

At the level of hadronic degrees of freedom, the linear response term $\Pi_{\mu\nu}(p)$ can be
written as
\begin{equation}
\Pi_{\alpha\beta}(p)F^{\alpha\beta} =
i\int d^4x  e^{ip\cdot x}\langle 0\,| \eta(x)
\left\{ -i \int d^4y A_\mu(y) J^\mu(y)\right\}
\bar{\eta}(0)\,|\,0\rangle.
\end{equation}
According to the basic assumption of current-hadron duality in the
QCD sum rules  approach \cite{Shifman79}, we insert  a complete
series of intermediate states satisfying the unitarity   principle
with the same quantum numbers as the current operator $\eta(x)$
 into the correlation function in
Eq.(4)  to obtain the hadronic representation. After isolating the
pole terms of the lowest pentaquark  states, we get the following result,
\begin{eqnarray}
\Pi_{\alpha\beta}(p)F^{\alpha\beta} & = &
-\int d^4x \int d^4y {d^4 k \over (2\pi)^4} {d^4 k^\prime \over (2\pi)^4}
\sum_{ ss^\prime}
\frac{1} {m^2_{\Theta^+} -k^2-i\epsilon}\frac{1} {m^2_{\Theta^+} -{k^\prime}^2-i\epsilon}
\nonumber \\ & &
e^{ip\cdot x} A_\mu(y)
\langle 0 | \eta(x) | ks \rangle
\langle ks | J^\mu(y) | k^\prime s^\prime \rangle
\langle k^\prime s^\prime | \bar{\eta}(0) | 0 \rangle+\cdots\,.
\label{phen2}
\end{eqnarray}
For the external electromagnetic field, it is convenient to use the fix-point gauge  $x_\mu A^\mu(x)=0$,
in this gauge, the electromagnetic potential is given by $
A_\mu(y)=-{1\over 2} F_{\mu\nu} y^\nu $.
The electromagnetic vertex in Eq.(5) can be parameterized   as
\begin{equation}
\langle ks | J^\mu(0) | k^\prime s^\prime \rangle =
\bar{u}(k,s) \left\{ \left[F_1(q^2)+ F_2(q^2)\right] \gamma^\mu
+\frac{(k+k^\prime)^\mu}{2m_{\Theta^+}}F_2(q^2)\right\}
u(k^\prime,s^\prime),
\end{equation}
where $q=k-k^\prime$. From the electromagnetic form factors  $F_1(q^2)$ and $ F_2(q^2)$ , we can obtain the
 magnetic moment  $\mu_{\Theta^+}$,
\begin{equation}
\mu_{\Theta^+}=\left\{ F_1(0)+F_2(0)\right\} \frac{e_{\Theta^+}}{2m_{\Theta^+}}.
\end{equation}
The linear response term $\Pi_{\mu\nu}(p)$ in the weak external electromagnetic field $F_{\alpha\beta}$  has
three different Dirac tensor structures,
\begin{equation}
\Pi_{\mu\nu}(p) =\Pi (p) \left\{\sigma_{\mu\nu} {\hat p} +{\hat p}\sigma_{\mu\nu}\right\}
+\Pi_1(p) i\left\{p_{\mu}\gamma_{\nu}-p_{\nu}\gamma_{\mu}\right\}{\hat p}
+\Pi_2(p) \sigma_{\mu\nu}  \, .
\end{equation}
The first structure has an odd number of $\gamma$-matrix and conserves chirality, the second and third have even
number of $\gamma$-matrixes and violate chirality. In the original QCD sum rules analysis of the nucleon magnetic
 moments \cite{Ioffe84,Balitsky83}, the interval of dimensions (of the condensates) for the odd
structure is larger than the interval of dimensions for the even structures, one may expect a better accuracy of the results
obtained from the sum rules at the odd structure. In this article, the spin of the pentaquark state $\Theta^+(1540)$ is supposed
to be $\frac{1}{2}$, just like the nucleon,
we can choose  the first Dirac tensor structure $\left\{\sigma_{\mu\nu} {\hat p} +{\hat p}\sigma_{\mu\nu}\right\}$.

After  performing  the integrals of the variables $x$, $y$, $k$ and $k^\prime$ in Eq.(5), and taking
the imaginary part, we obtain the phenomenological  spectral density,
\begin{equation}
\mbox{Im} \Pi (s) =
{1\over 4} \{F_1(0)+F_2(0)\} f_0^2
\delta^\prime (s-m_{\Theta^+}^2)
+ C_{subtract} \delta (s-m_{\Theta^+}^2)
+\cdots\, ,
\end{equation}
where the first term corresponds to the magnetic moment $\mu_{\Theta^+}$, and is
of double-pole. The second term comes from the electromagnetic transitions between the pentaquark state $\Theta^+(1540)$
 and the excited states,  and is of single-pole. We have not shown the contributions from the higher
resonances and  continuum states explicitly for simplicity.

In the following, we perform the operator product expansion in the weak external electromagnetic field
 to obtain the spectral representation at the level of
quark and gluon degrees of freedom.
  The  calculation of  operator product expansion in the  deep Euclidean space-time region is
  straightforward and tedious, here technical details are neglected for simplicity,
  once  the analytical  results are obtained,
  then we can express the correlation functions at the level of quark-gluon
degrees of freedom into the following form through dispersion
relation,
  \begin{eqnarray}
  \Pi(P^2)= \frac{1}{\pi}\int_{m_s^2}^{s_0}ds
  \frac{{\rm Im}[A(-s)]}{s+P^2}+
  B(P^2)+\cdots\, ,
  \end{eqnarray}
where
\begin{eqnarray}
\frac{{\rm Im}[A(-s)]}{\pi}&=&
-\frac{3\left[14e_u+3e_d+2e_s\right]s^4}{2^{15}6!4!\pi^8}+\frac{\left[-(e_u+2e_d)m_s\chi \langle
\bar{q}q\rangle+2e_s m_s\chi \langle\bar{s}s \rangle \right] s^3}{2^{11}5!4!\pi^6} \nonumber \\
&-& \left\{ (e_u-2e_d)\chi \langle \bar{q}q\rangle^2-\frac{e_u+2e_d+6e_s}{3}\chi \langle \bar{q}q\rangle \langle \bar{s}s\rangle
+\frac{(3e_d+14e_u)m_s\langle \bar{s}s\rangle}{8\pi^2 } \right. \nonumber \\
&-&\left.  \frac{5(4e_u+e_d)m_s\langle \bar{q}q\rangle}{4\pi^2} \right\} \frac{ s^2}{2^{12}4!\pi^4}\nonumber\\
&-&\left\{ (4e_u+e_d)\left[ \langle \bar{q}q\rangle^2 +\langle \bar{q}q\rangle\langle \bar{s}s\rangle\right]
+\frac{8e_s\langle \bar{q}q\rangle^2}{5} \right\}\frac{5s}{2^{12}3^2\pi^4}  \nonumber\\
&+&\left\{ 3(e_u-2e_d)m_s\chi \left[2\langle \bar{q}q\rangle^3 -\langle \bar{q}q\rangle^2\langle \bar{s}s\rangle\right]
+16e_sm_s\chi\langle \bar{q}q\rangle^2\langle \bar{s}s\rangle\right\}\frac{1}{2^{11}3^2\pi^2} \nonumber,
\end{eqnarray}
\begin{eqnarray}
B(P^2)&=&\left\{ \frac{e_u+2e_d}{3}\chi \langle \bar{q}q\rangle^4
-(e_u-2e_d-2e_s)\chi \langle \bar{q}q\rangle^3\langle \bar{s}s\rangle
+\frac{14e_u+3e_d}{4\pi^2}m_s\langle \bar{q}q\rangle^3 \right. \nonumber\\
&-&\left. \frac{5}{8\pi^2}(4e_u+e_d)m_s \langle \bar{q}q\rangle^2\langle \bar{s}s\rangle\right\} \frac{1}{2^7 3^2 \pi^2 P^2} \nonumber \\
&+&\left\{14e_u \langle \bar{q}q\rangle^3\langle \bar{s}s\rangle+3e_d\langle \bar{q}q\rangle^3\langle \bar{s}s\rangle
+2e_s\langle \bar{q}q\rangle^4\right\}\frac{1}{2^7 3^3 P^4} .
\end{eqnarray}
The presence of the external electromagnetic field $F_{\mu\nu}$ induces
three new vacuum condensates i.e. the vacuum susceptibilities in the QCD vacuum \cite{Ioffe84,Balitsky83},
\begin{eqnarray}
\langle  {\overline  q} \sigma_{\mu\nu} q  \rangle_{F_{\mu\nu}} = e_q  \chi
F_{\mu\nu} \langle  {\overline  q}  q  \rangle \, , \nonumber \\
g_s \langle  \bar{q} G_{\mu\nu} q  \rangle_{F_{\mu\nu}}
= e_q  \kappa F_{\mu\nu}
\langle  {\overline  q}  q  \rangle \, , \nonumber\\
g_s \epsilon^{\mu\nu\lambda\sigma}
\langle  {\overline  q} \gamma_5  G_{\lambda\sigma} q |0 \rangle_{F_{\mu\nu}}
= i e_q  \xi F^{\mu\nu}
\langle  {\overline  q}  q  \rangle \, , \nonumber
\end{eqnarray}
where $e_q$ is the quark charge,
the $\chi$, $\kappa$ and $\xi$  are the  vacuum
susceptibilities. The values with  different theoretical
approaches are different from each other, for a short review,  one can see Ref.\cite{Wang02}.
Here we shall adopt the values
$\chi=-4.4\, \mbox{GeV}^{-2}$,
$\kappa =0.4$ and $\xi = -0.8$  \cite{Ioffe84,Balitsky83,Belyaev84}. In calculation,
we have neglected the terms which concern the mixed vacuum condensates    $ \langle
g_s\bar{q}\sigma  Gq \rangle$ and $ \langle
g_s\bar{s}\sigma  Gs \rangle$ as they are suppressed by large denominators;
we have also neglected the terms of the form
$e_q m_s {\rm log}(-x^2)$ due to the small current quark mass $m_s$.

The Borel transformation with respect to the variable $P^2$
is straight forward and easy,
 \begin{equation}
\Pi (M^2) \equiv \lim_{n,P^2 \rightarrow \infty} \frac{1}{\Gamma(n)}
(P^2)^{n} \left( - \frac{d}{d P^2} \right)^n \Pi (P^2),
\end{equation}
  with the  Borel parameter  $M^2
= P^2/n$ kept fixed in the limit. Finally we obtain the sum rules for the form factors   $F_1(0)$ and $F_2(0)$,
\begin{eqnarray}
-\frac{1}{4} \left\{F_1(0)+F_2(0)\right\}\frac{1+CM^2}{M^4}f_0^2e^{-\frac{m^2_{\Theta^+}}{M^2}}=\frac{1}{M^2}\int_{m_s^2}^{s_0}
ds  \frac{{\rm Im}[A(-s)]}{\pi}e^{-\frac{s}{M^2}}+
  B(M^2),
\end{eqnarray}
where $m_s$ is the strange quark
mass and $s_0$ is the
threshold parameter used to subtract the contributions from the higher
resonances and  continuum states. Here the $C$ denotes the contributions from
 the single-pole terms and is proportional to the quantity $C_{subtract}$ in Eq.(9).
  We have no knowledge about the  electromagnetic transitions between the pentaquark state $\Theta^+(1540)$
 and the excited states, the $C$ can be taken as a free parameter, we  choose the suitable values for $C$ to
  eliminate the contaminations from  the single-pole terms to obtain the reliable  sum rules. In practical calculations,
  it can be fitted to give stable sum rules with respect to
 variation  of the Borel parameter $M^2$ in a suitable interval.
 Furthermore, from the correlation function $\Pi_0(p)$ in Eq.(1), we can obtain the sum rules for the coupling constant $f_0$ \cite{zhu03},
\begin{equation}
f^2_0e^{-{m_{\Theta^+}^2\over M^2}}=\int_{m_s^2}^{s_0} ds e^{-{s\over M^2}}
\rho_0 (s),
\end{equation}
where
\begin{eqnarray}
 \rho_0={3
s^5\over  4^8 7! \pi^8}+{s^2\over 1536 \pi^4} \left[ {5\over 12}
\langle \bar q q\rangle^2  +{11\over 24}\langle \bar
qq\rangle\langle \bar ss\rangle \right ] +\left[{7\over
432}\langle \bar q q\rangle^3 \langle \bar s s\rangle +{1\over
864}\langle \bar q q\rangle^4\right] \delta (s) . \nonumber
\end{eqnarray}
From above equations, we can obtain
\begin{eqnarray}
F_1(0)+F_2(0) =-4 \frac{ M^2\int_{m_s^2}^{s_0}
ds  \frac{{\rm Im}[A(-s)]}{\pi}e^{-\frac{s}{M^2}}+
  M^4 B(M^2)}{\left\{1+CM^2\right\}\int_{m_s^2}^{s_0}ds e^{-{s\over M^2}}
\rho_0 (s)}.
\end{eqnarray}

\section{Numerical Results}
The parameters are taken as $\langle \bar{s}s \rangle=0.8\langle
\bar{u}u \rangle$, $\langle \bar{u}u \rangle=\langle \bar{d}d
\rangle=\langle \bar{q}q \rangle=(-219 MeV)^3$, $\chi=-4.4GeV^{-2}$,
 $m_u=m_d=0$ and $m_s=150MeV$. Here we have neglected the uncertainties about
 the vacuum condensates, small variations of those condensates will not
 lead to larger changes about   the numerical
 values.     In calculation, we have also neglected
  the terms which concern the mixed  vacuum condensates    $ \langle
g_s\bar{q}\sigma Gq \rangle$ and $ \langle
g_s\bar{s}\sigma Gs \rangle$ as they are suppressed by large denominators;
we have also neglected the terms of the form
$e_q m_s {\rm log}(-x^2)$ due to the small current quark mass $m_s$.
 The threshold parameter $s_0$ is chosen to  vary between $(3.7-4.3) GeV^2$ to avoid possible contaminations from
  higher resonances and continuum states. In the region $M^2=(1.5-5.5)GeV^2$, the sum rules for $F_1(0)+F_2(0)$ are almost independent of
 the Borel parameter $M^2$.  For $s_0=(3.7-4.3) GeV^2$, we obtain the values
 \begin{eqnarray}
 F_1(0)+F_2(0)&=&0.40\pm 0.03 \, , \nonumber\\
 \mu_{\Theta^+}&=& (0.40\pm 0.03) \frac{e_{\Theta^+}}{2m_{\Theta^+}}\, , \nonumber \\
  &=& (0.24\pm0.02)\mu_N,
  \end{eqnarray}
where the $\mu_N$ is the nucleon  magneton.
 From the Table 1,  we can see that although  the numerical values for
 the magnetic moment  $ \mu_{\Theta^+}$  vary with theoretical
 approaches, they are small in general;  our numerical results
are consistent with most of the existing values of theoretical estimations.
We have studied the magnetic moment $\mu_{\Theta^+}$ from the first principle of QCD,
although the uncertainties  of the condensates, the neglect of the higher
dimension condensates,  the lack of perturbative QCD corrections,
etc, will  result in errors, the predictions are robust, or qualitative at least.

In Ref.\cite{Nam04M}, the anomalous magnetic moment $\kappa_{\Theta^+}$ of the
pentaquark state $\Theta^+(1540)$ was first estimated in the photo-production process,
for example, $\kappa_{\Theta^+} = - 0.7{e_{\Theta^+}\over
2m_{\Theta^+}}$ for
$J^P={1\over 2}^+$ state and $\kappa_{\Theta^+} = - 0.2{e_{\Theta^+}\over
2m_{\Theta^+}}$  for $J^P={1\over 2}^-$ state using the Jaffe-Wilczek's diquark picture.
In Ref.\cite{Zhao04M}, the magnetic moment $\mu_{\Theta^+}$ was
calculated in a quark model, $\mu_{\Theta^+}=0.13{e_{\Theta^+}\over
2m_{\Theta^+}}$ for positive parity state using Jaffe-Wilczek's diquark picture,
 and $\mu_{\Theta^+}={e_{\Theta^+}\over
6m_{s}}$ for negative parity state with the quark clusters $u\bar{s}$ and $udd$.

In the chiral quark soliton model, one obtain the magnetic moment $\mu_{\Theta^+}=(0.2-0.3)\mu_N$ in the
chiral limit with the parameters calculated in a model-dependent way \cite{Kim04M}, while
with the parameters adjusted in a model-independent way, one obtain the magnetic moment $\mu_{\Theta^+}$
which strongly depends  on the pion-nucleon sigma term $\Sigma_{\pi N}$,
 $\mu_{\Theta^+}=-1.19\mu_N$  for $\Sigma_{\pi N}=45MeV$, $ \mu_{\Theta^+}=-0.33\mu_N$
 for $\Sigma_{\pi N}=75MeV$ \cite{Goeke04M}.

In Ref.\cite{Zhu04M}, the authors calculate the magnetic moment $\mu_{\Theta^+}$
 in four quark models within the framework of quantum mechanics, and obtain the values $\mu_{\Theta^+}=0.08\mu_N$ for
Jaffe-Wilczek's diquark(scalar)-diquark(scalar)-antiquark model; $\mu_{\Theta^+}=0.23\mu_N$
for Shuryak-Zahed's diquark(scalar)-diquark(tensor)-antiquark model; $\mu_{\Theta^+}=0.19\mu_N$
for Karliner-Lipkin's diquark-triquark model; $\mu_{\Theta^+}=0.37\mu_N$ for Strottman's model.
 In Ref.\cite{Bijker04M}, the authors calculate the magnetic moment $\mu_{\Theta^+}$
  in the constituent quark model, and obtain
  $\mu_{\Theta^+}=0.38\mu_N$ for $ J^P={\frac{1}{2}}^-$ state,
 $\mu_{\Theta^+}=0.09\mu_N$ for $ J^P={\frac{1}{2}}^+$ state.
In Ref.\cite{Inoue04M}, the magnetic moment $\mu_{\Theta^+}$
is calculated in the naive additive quark model, $\mu_{\Theta^+}=0.4\mu_N$ for $J^P={\frac{1}{2}}^-$ state.
  In Ref.\cite{Hong04M}, the authors calculate the magnetic moment $\mu_{\Theta^+}$ in chiral effective theory with Jaffe-Wilczek's
   diquark model, and obtain $\mu_{\Theta^+}=(0.56-0.71)\mu_N$.
In Ref.\cite{Delgado04M}, the magnetic moment $\mu_{\Theta^+}$  is calculated in the diquark-triquark configuration in cluster quark
model with color-magnetic spin-spin interactions.

In Ref.\cite{Huang04M}, the magnetic moment $\mu_{\Theta^+}$  is calculated
with  the same interpolating current in the framework of light-cone QCD sum rules.
 In the light-cone QCD sum rules approach, the uncertainties  of the photon light-cone
distribution amplitudes, keeping only the lowest-twist few terms
of two particles, the uncertainties  of the condensates, the neglect
of the higher dimension condensates, the lack of perturbative QCD
corrections, etc, can lead to errors in predictions and impair  the prediction power.
Our results are consistent with the small values obtained in Ref.\cite{Huang04M}.
In Ref.\cite{Wang052} and Ref.\cite{Wang053}, we take the diquark-diquark-antiquark type interpolating current
to calculate the magnetic moment  $\mu_{\Theta^+}$ in different approaches (i.e. the QCD sum rules in the external
field and the light-cone QCD sum rules), and obtain $\mu_{\Theta^+}=-(0.134\pm0.006)\mu_N$ and
$\mu_{\Theta^+}=-(0.49\pm0.06)\mu_N$, respectively. Due to the special interpolating
current, only the interactions of  the $s$ quark with
the external electromagnetic field  have contributions to the
tensor structure $\sigma_{\mu\nu} {\hat p} +{\hat
p}\sigma_{\mu\nu}$ for the QCD sum rules in the external
field, and the tensor structure $\varepsilon_{\mu\nu\alpha\beta}\gamma_5\gamma^\mu\varepsilon^\nu
q^\alpha p^\beta$ for the light-cone QCD sum rules, respectively,
 which is significantly different from the
results obtained in Ref.\cite{Huang04M} and the present work, where
 all the $u$, $d$ and $s$ quarks have contributions.
 If we take the same interpolating current,
 the magnetic moment $\mu_{\Theta^+}$ obtained from the QCD sum rules in the external
  field are more stable than the corresponding one from the light-cone QCD  sum rules with respect
 to the variations of the Borel parameter $M^2$. The analysis of the magnetic moment $\mu_{\Theta^+}$
 with  other interpolating currents in the framework of QCD sum rules approach will be our next work \cite{sugiyama04,eide04}.

When  the experimental measurement of the
  magnetic moment of the pentaquark state $\Theta^+(1540)$
  is possible in the near future, we might  be able  to test the theoretical
   predictions of the magnetic moment $\mu_{\Theta^+}$ and  select
the preferred quark configurations and QCD sum rules.

  \begin{table}[ht]
         \caption{\label{Tablechi}The values of $\mu_{\Theta^+}$ (in unit of $\mu_N$)}
         \begin{center}
         \begin{tabular}{c||c}
         \hline\hline
         Reference          & $\mu_{\Theta^+}$ \\
                            &  $(\mu_N)$  \\ \hline\hline
      \cite{Nam04M}        & 0.2$\sim$0.5         \\ \hline
       \cite{Zhao04M}     &   0.08 $\sim$ 0.6        \\ \hline
         \cite{Kim04M} & 0.2$\sim$0.3     \\ \hline
      \cite{Goeke04M}             &-1.19 or -0.33         \\ \hline
      \cite{Zhu04M}              & 0.08 or 0.23 or 0.19 or 0.37           \\ \hline
       \cite{Bijker04M}             & 0.38          \\ \hline
      \cite{Inoue04M}              & 0.4          \\ \hline
          \cite{Hong04M}             &0.71 or 0.56          \\ \hline
      \cite{Delgado04M}            &0.362          \\ \hline
       \cite{Huang04M}         &   0.12 $\pm$ 0.06        \\ \hline
       \cite{Wang052}         &   -(0.134$\pm$ 0.006)        \\ \hline
       \cite{Wang053}         &   -(0.49$\pm$ 0.06)        \\ \hline
                   This Work          &0.24$\pm$0.02           \\ \hline\hline
         \end{tabular}
         \end{center}
         \end{table}

\section{Conclusion }

In summary, we have calculated  the magnetic moment of the
 pentaquark state $\Theta^+(1540)$ with the  QCD sum
rules  approach in the weak external electromagnetic field. The numerical results
are consistent with most of the existing values of theoretical estimations,
$ \mu_{\Theta^+}= (0.40\pm 0.03) \frac{e_{\Theta^+}}{2m_{\Theta^+}}
  = (0.24\pm0.02)\mu_N$ .
The  magnetic moments of the
baryons are  fundamental parameters as their masses, which have
copious  information  about the underlying quark structures,
different substructures can lead to
 very different results. The width of the pentaquark state $\Theta^+(1540)$ is so
narrow, the
 small magnetic moment $\mu_{\Theta^+}$ may be extracted from the electro- or photo-production
experiments eventually in the near future, which may be used to
distinguish the preferred quark configurations and QCD sum rules from
various theoretical models, obtain more insight into the relevant
degrees of freedom and deepen our understanding about  the
underlying dynamics that determines the properties of the exotic
pentaquark states.

\section*{Acknowledgment}
This  work is supported by National Natural Science Foundation,
Grant Number 10405009,  and Key Program Foundation of NCEPU. The
authors are indebted to Dr. J.He (IHEP), Dr. X.B.Huang (PKU) and Dr. L.Li (GSCAS)
for numerous help, without them, the work would not be finished.

\end{document}